\documentstyle[12pt]{article}
\def\tder{\partial_{\theta}}

\def\zcl1{{\cal CZ}^1_3}

\def\0m{M_{\bar 0}(3,{\bf C})}
\def\1m{M_{\bar 1}(3,{\bf C})}
\def\2m{M_{\bar 2}(3,{\bf C})}

\def\noi{\noindent}
\def\TCA{{\cal T}_q(N)}
\def\three{{\cal T}_q^0(3)}
\def\commut{[f_1,f_2,f_3]}
\def\123{f_1\,f_2\,f_3}
\def\231{f_2\,f_3\,f_1}
\def\312{f_3\,f_1\,f_2}
\begin{document}
\bibliographystyle{unsrt}
%%%%%%%%%%%%%%%%%%%%%%%%%%%%%%%%%%%%%%%%%%%%%%%%%%%%%%%%%%%%%%%%%%%%%%
%%%%%%%%%%%%%% TITLEPAGE %%%%%%%%%%%%%%%%%%%%%%%%%%%%%%%%%%%%%%%%%%%%%

\title{\huge\rm $Z_3$-graded analogues of Clifford algebras and
generalization of supersymmetry}

\author{Viktor Abramov\\*[-.5ex]{\normalsize\it Institute of Pure
Mathematics, University of Tartu,}\\*[-.5ex]{\normalsize\it Vanemuise 46,
EE2400 Tartu, ESTONIA}} 
\date{}
\maketitle
%%%%%%%%%%%%%%%%%%%%%%%%%%%%%%%%%%%%%%%%%%%%%%%%%%%%%%%%%%%%%%%%%%%%%%
{\abstract We define and study the ternary analogues of
Clifford algebras. It is proved that the ternary Clifford algebra with $N$
generators is isomorphic to the subalgebra of the elements of grade zero
of the ternary Clifford algebra with $N+1$ generators. In the case $N=3$
the ternary commutator of cubic matrices induced by the ternary commutator
of the elements of grade zero is derived. We apply the
ternary Clifford algebra with one generator to construct the
generalization of the simplest algebra of supersymmetries.}
%%%%%%%%%%%%%%%%%%%%%%%%%%%%%%%%%%%%%%%%%%%%%%%%%%%%%%%%%%%%%%%%%%%%%%%
\section{Introduction}
For the last few years there have appeared the generalizations 
of supersymmetry. It is well known that from mathematical point of view
the concept of supersymmetry is 
based on the $Z_2$-graded structures. The generalizations of supersymmetry
are usually constructed with the help of $Z_n$-graded structures, where
$n=3,4,\ldots$. Therefore the $Z_n$-graded generalizations of the
Grassmann and Clifford algebras may be useful in constructing the
generalized supersymmetries.

In this paper we define and study the analogues of Clifford algebras
which have a natural $Z_3$-grading. Although the generators of these
algebras are subjected to the {\it binary}
commutation relations (\ref{commut-relat}) the analogy with the classical
Clifford algebras appears when we derive the relations (\ref{Clifford})
involving {\it three} generators. This is the reason why we use the term
"ternary Clifford algebra" (TCA).

We show that the generalizations of Clifford algebras proposed in this paper
have the properties which are very similar to those of the classical Clifford
algebras. Particularly we prove that the TCA with $N$ generators is
isomorphic to the subalgebra of the elements of grade zero of the TCA with
$N+1$ generators. In section 3 we explore the structure of the TCA with
respect to ternary commutator. It is proved that in the case $N=3$ the
space spanned by the monomials of grade zero is closed with respect to
ternary commutator and this fact is used to derive the formula for ternary
commutator of cubic matrices. In section 4 we propose the $Z_3$-graded
version of the
ternary commutator and we use it to construct the generalization of the
simplest algebra of supersymmetries. This algebra is closely related
to the concept of fractional supersymmetry \cite{Colatto}-\cite{Mohammedi}.
%%%%%%%%%%%%%%%%%%%%%%%%%%%%%%%%%%%%%%%%%%%%%%%%%%%%%%%%%%%%%%%%%%%%%%%
\vskip .5cm
\section{Ternary Clifford algebra (TCA)}
{\bf 1. Definition of TCA}. Let $q=exp(2\pi/3)$ be the cube root of unit
and $G=\Vert q_{ij}\Vert$ 
be the $N\times N$-matrix whose entries are defined as follows
\begin{equation}
q_{ij}=\left\{\begin{array}{ll}
  1,   & \mbox{$i=j$}\\
  q,   & \mbox{$i>j$}\\
  q^2, & \mbox{$i<j$}
 \end{array}
\right.
\label{q-definition}
\end{equation}
It is clear that $G$ is a Hermitian matrix, i.e. ${\bar G}^t=G$. For each
triple of indices
$1\leq j,k,l\leq N$ such that 
\begin{equation}
\vert j-k\vert+\vert
k-l\vert+\vert j-l\vert\not=0,
\label{indices}
\end{equation}
the entries of the matrix $G$ satisfy
the identity
\begin{equation}
1+q_{kl}+q_{jk}+q_{jl}\,q_{kl}+q_{jl}\,q_{jk}+q_{kl}\,q_{jk}\,q_{jl}=0,
\label{identity}
\end{equation}
which will play an essential role in what follows.

The
{\it ternary Clifford algebra} (TCA) is an associative algebra over the
field ${\bf C}$ with the identity element 1 generated by the symbols
$Q_1,\,Q_2,\,\ldots,\,Q_N$ such that 
\begin{equation}
Q_i\,Q_j=q_{ij}\;Q_j\,Q_i,
\label{commut-relat}
\end{equation}
and
\begin{equation}
Q_i^3=1,\quad i=1,2,\ldots,N.
\label{cubic-relations}
\end{equation}
Let us denote the above defined TCA by $\TCA$. It is clear that replacing
the relations (\ref{cubic-relations}) in the above definition by
$Q_i^3=-1,\;i=1,2,\ldots,N$ one does not change the structure of the
algebra. In section 4 we shall demonstrate that TCA with
$Q_i^3=-1,\;i=1,2,\ldots,N$ is closely related to 
the fractional supersymmetry and we shall denote it by ${\cal T}_q^{-}(N)$. 

Let us define the analogue of the Kronecker symbol by
the formula
\begin{equation}
\delta_{ijk}=\sum_l^N\,\delta^l_i\,
                                   \delta^l_j\,\delta^l_k.
\label{delta}
\end{equation}
From (\ref{identity}) and (\ref{commut-relat}) it
follows then that the generators of the TCA satisfy the ternary relations
\begin{equation}
\{Q_i,Q_j,Q_k\}=6\,\delta_{ijk},
\label{Clifford}
\end{equation}
where the braces at the left-hand side stand for the sum of products of all
permutations of the generators $Q_i,Q_j,Q_k$ which can be called {\it ternary
anticommutator.} 
The relations (\ref{Clifford}) have the form which is
similar to that of the commutation relations usually assumed as a basis
for the definition of the classical Clifford algebras. Particularly from
(\ref{Clifford}) it follows that the generators of the TCA are ternary
anticommutative that is if the indices $i,j,k$ satisfy the
condition (\ref{indices}) then $\{Q_i,Q_j,Q_k\}=0.$ 

\vskip .3cm
\noi
{\bf 2. General structure and dimension.} In order to give a more precise
description of the structure of the TCA 
it is useful to introduce the generators with bars on subscripts. Let us
define $Q_{\bar i}=Q^2_i$. From (\ref{commut-relat}) it follows then that
generators with bars on subscripts satisfy the relations
\begin{equation}
Q_i\,Q_{\bar k}=q_{i{\bar k}}\;Q_{\bar k}\,Q_i,\quad
    Q_{\bar i}\,Q_k=q_{{\bar i}k}\;Q_k\,Q_{\bar i},\quad   
Q_{\bar i}\,Q_{\bar k}=q_{{\bar i}{\bar k}}\;Q_{\bar k}\,Q_{\bar i}, 
\label{add-commut-relat}
\end{equation}
where $q_{i{\bar k}}=q_{{\bar i}k}={\bar q}_{ik}$ and $q_{{\bar i}{\bar
k}}=q_{ik}$. In order to combine the above commutation relations and the
relations (\ref{commut-relat}) into a single formula 
we shall use the subscripts $A,B,C,\ldots$ running both
sets ${\cal N}=\{1,2,\ldots,N\}$ and 
$\bar {\cal N}=\{\bar 1,\bar 2,\ldots,\bar N\}$. We 
define the matrix 
${\cal G}=\Vert q_{AB}\Vert$ to be 
$$
{\cal G}=\left\Vert\matrix{q_{ik} & q_{i\bar k}\cr
q_{{\bar i}k} & q_{{\bar i}{\bar k}}\cr}\right\Vert,
$$
where $i,k$ run from 1 to $N$. Now the commutation relations
(\ref{commut-relat}) and (\ref{add-commut-relat}) can be combined into a
single formula
\begin{equation}
Q_A\,Q_B=q_{AB}\;Q_B\,Q_A.
\end{equation}
It is easy to see that the matrix ${\cal G}$ which is an extension of $G$
is a Hermitian matrix.

The structure of the TCA becomes more transparent if we use the notations
which are similar to the Kostant ones usually used in classical Grassmann
and Clifford algebras.
Let $I=\{i_1,i_2,\ldots,i_k\},\;J=\{j_1,j_2,\ldots,j_l\},\;
i_1<i_2<\ldots<i_k,\,j_1<j_2<\ldots<j_l$ be two 
subsets of ${\cal N}$ such that $I\cap J=\emptyset$. Let us denote 
$Q_I=Q_{i_1}\,Q_{i_2}\ldots Q_{i_k}$ and $Q_{\bar J}=Q_{\bar j_1}\,Q_{\bar
j_2}\ldots Q_{\bar j_l}$. Then any element $f$ of $\TCA$ can
be expressed in terms of the monomials $Q_I\,Q_{\bar J}$ as follows
\begin{equation}
f=\sum_{I,J} {\alpha}_{I\bar J}\;Q_I\,Q_{\bar J},
\end{equation}
where ${\alpha}_{I\bar J}$ are complex numbers and the sum is taken over all
possible pairs of subsets $(I,J)$. Thus the monomials $Q_I\,Q_{\bar J}$
constitute the basis of the vector space underlying the algebra $\TCA$.
Counting of these monomials yields the dimension of the vector space
associated to the TCA which is $3^N$. This result points to the analogy
with the classical Clifford algebra since the dimension of the vector
space of the classical Clifford algebra with $n$ generators is $2^n$.

\vskip .3cm
\noi
{\bf 3. ${\bf Z_3}$-grading and tensor product.} TCA has a natural
$Z_3$-grading. There are two ways to define the $Z_3$-grading of TCA which
are conjugate to each other. The first one is to associate grade 0 to the
identity element and grade 1 to the generators $Q_i$. The second one is to
associate as before the grade 0 to the identity element and grade 2 to the
generators $Q_i$. Let us denote the grade of an element $f$ by $gr(f)$. As
usual we define the grade of any monomial $Q_I$ as the sum of the grades
of its generators modulo 3. Clearly that $gr(Q_{\bar i})=2$ in the first
case and $gr(Q_{\bar i})=1$ in the second. We shall use the grading
defined by the first way though the second grading could be used equally
well. Then the algebra $\TCA$ splits into the direct sum of its subspaces
\begin{equation}
\TCA=T_q^{0}(N)+T_q^1(N)+T_q^2(N),
\end{equation}
each consisting of the elements respectively of grade 0,1 and 2. The subspace
$T_q^{0}(N)$ of the elements of grade 0 is a subalgebra of
$\TCA$. It can be shown that the dimension of this subalgebra is $3^{N-1}$. 

In analogy with superstructures we define the $Z_3$-graded $q$-tensor
product of two TCA. Given two algebras 
${\cal T}_q(N),{\cal T}_q(N')$  
generated respectively by $Q_1,Q_2,\ldots,Q_N$ and 
$Q'_1,Q'_2,\ldots,Q'_{N'}$ we form the {\it $Z_3$-graded $q$-tensor product}
${\cal T}_q(N)\otimes_q {\cal T}_q(N')$ which is the tensor
product of the underlying vector spaces equipped with the multiplication
\begin{equation}
(f\otimes_q f')\,(h\otimes_q h')=q^{gr(f')\,gr(h)}\;ff'\otimes_q hh'.
\end{equation}
If we identify $Q_i\equiv Q_i\otimes_q 1$ and denote $Q_{N+1}=1\otimes_q
Q'_1,\ldots,Q_{N+N'}=1\otimes_q Q'_{N'}$ then 
${\cal T}_q(N)\otimes_q {\cal T}_q(N')\cong {\cal T}_q(N+N').$ 

\vskip .3cm
\noi
{\bf 4. Isomorphism ${\bf {\cal T}_q(N)\cong T^0_q(N+1)}$.}
It is well known that if ${Cl}(n),$\ \ ${Cl}(n+1)$ are
two classical Clifford algebras over the field ${\bf C}$ generated by
$\gamma_1,\gamma_2,\ldots,\gamma_{n+1}$ such that 
$\{\gamma_i,\gamma_j\}=2\,\delta_{ij}$ then there is the
isomorphism 
\begin{equation}
r:\,{Cl}(n)\to {Cl}^0 (n+1),
\label{isomorphism}
\end{equation}
where 
${Cl}^0(n+1)$ means the even subalgebra of ${Cl}(n+1)$. This
isomorphism is defined by $r(a_0+a_1)=a_0+i\,a_1\,\gamma_{n+1}$ where
$a_0, a_1$ are respectively the even and odd parts of an element. 
It is worth mentioning
that this isomorphism plays an essential role in the
theory of Dirac operator \cite{Trautman}.

It turns out that a similar isomorphism can be constructed for the ternary
Clifford algebras as well. Given two TCA's ${\cal T}_q(N)$ and ${\cal
T}_q(N+1)$ 
we define the mapping $\zeta:\,{\cal T}_q(N)\to T^0_q(N+1)$
by the formula 
\begin{equation}
\zeta(f)=f_0+p\,f_1\,Q_{\overline{N+1}}+
                       p^{-1}\,{\bar q}\,f_2\,Q_{N+1}, 
\end{equation}
where $p^3=1$ and $f_0,f_1,f_2$ are the parts of an element $f$
respectively of the grades $0,1,2$. This mapping is the isomorphism of the
algebras that can be proved by means of the relations
$$
Q_{N+1}\,f_k=q^k\,f_k\,Q_{N+1},\quad
    Q_{\overline{N+1}}\,f_k={\bar q}^k\,f_k\,Q_{\overline{N+1}},
                   \quad k=0,1,2.
$$

\vskip .3cm
\noi
{\bf 5. Involution.} Let $\TCA$ be the ternary Clifford algebra generated by
$Q_1,Q_2,\ldots,Q_N$. One can define an involution on the algebra $\TCA$.
Since any element $f$ of the algebra $\TCA$ can be expressed in terms of
the monomials $Q_I\,Q_{\bar J},\;I=(i_1,i_2,\ldots,i_k),\;
J=(j_1,j_2,\ldots,j_l),\;I\cap J=\emptyset$ it is sufficient to define the
involution for these monomials and then extend it by linearity to an
arbitrary element of the TCA. Let us define the mapping $*:\TCA\to \TCA$
by the formula
\begin{equation}
(\alpha\,Q_I\,Q_{\bar J})^{*}=
   {\bar \alpha}\,Q_{j_{l}}Q_{j_{l-1}}\ldots Q_{j_{1}}\,
        Q_{{\overline i_{k}}}Q_{{\overline i_{k-1}}}\ldots 
                                           Q_{{\overline i_{1}}},
\label{involution}
\end{equation}
where $\alpha\in {\bf C}$. Using the commutations relations
(\ref{commut-relat}) of TCA one can put the right-hand side of the above
formula into the form
\begin{equation}
(\alpha\,Q_I\,Q_{\bar J})^{*}=
   {\bar \alpha}\,q^m\,Q_J\,Q_{\overline I},
\end{equation}
where $m={1/2}\,(k(k-1)+l(l-1))$. The mapping defined by
(\ref{involution}) enjoys the following properties:
\begin{equation}
(\alpha\,f_1+\beta\,f_2)^{*}=
         {\bar \alpha}\,f^{*}_1+{\bar \beta}\,f_2^{*},\quad
(f_1\,f_2)^{*}=f^{*}_2\,f^{*}_1,\quad (f^{*})^{*}=f
\label{properties}
\end{equation}
where $\alpha,\beta\in {\bf C},\;f,f_1,f_2\in \TCA$. From
(\ref{properties}) it follows that the mapping $*:\TCA\to \TCA$ is an
involution of the algebra $\TCA$. This involution does not change the
grades of the elements of the sublagebra $T^0_q(N)$ and
transforms the elements of grade 1 to the elements of grade 2 and vise
verse. 

\section{Ternary commutator and the Lie structure of TCA.}
{\bf 1. Ternary commutator and its group properties.} It is well known
that the classical Clifford algebras can be used to construct the important
examples of Lie algebras. The construction of these Lie algebras is based
on the notion of commutator. We shall use the notion
of the ternary commutator, which can be viewed as a generalization of the
ordinary commutator, to study the similar structures of the ternary
Clifford algebras. Let $\TCA$ be the ternary Clifford algebra and
$f_1,f_2,f_3\in \TCA$. We define the ternary $p$-commutator by the formula
\begin{eqnarray}
\lefteqn{\commut_p=\123+p\;\231+{\bar
        p}\;\312+}\nonumber\qquad\qquad\qquad\\ & & 
                  \quad\qquad f_3\,f_2\,f_1+p\;f_1\,f_3\,f_2+
                                       {\bar p}\;f_2\,f_1\,f_3,
\label{p-commutator}
\end{eqnarray}
where $p^3=1,\,p\not=1$. The above formula can be given a more shorter
form if we define $f_1*f_2*f_3=f_1\,f_2\,f_3+(f_1\,f_2\,f_3)^{*}$. Then 
(\ref{p-commutator}) can be written in the form
\begin{equation}
\commut_p=f_1*f_2*f_3+p\;f_2*f_3*f_1+{\bar p}\;f_3*f_1*f_2.
\end{equation}

It is easy to see that just as in the case of the classical
commutator the ternary $p$-commutator equals zero as soon as one of
its arguments is the identity element, i.e.
\begin{equation}
[1,f,h]_p=[f,1,h]_p=[f,h,1]_p=0,\qquad f,h\in {\cal T}_q(N).
\end{equation}

It is obvious that there are two choices in the
above definition: $p=q$ and $p={\overline q}$. Using involution
(\ref{involution}) we define the conjugate
ternary commutator by the formula
\begin{equation}
\commut^{\dagger}_p=[f^{*}_1,f^{*}_2,f^{*}_3]^{*}_p.
\label{con-of-commutator}
\end{equation}
The straightforward computation shows that
\begin{equation}
\commut^{\dagger}_p=[f_1,f_2,f_3]_{\bar p}.
\end{equation}

The important property of the classical binary commutator is its
anticommutativity, i.e. $[a_2,a_1]=-[a_1,a_2]$. Let $S_2$ be the
group of permutations of two elements and $w:S_2\to \{-1,1\}$ its
representation. Then the anticommutativity can be written in the form
\begin{equation}
[a_{\sigma(2)},a_{\sigma(1)}]=w(\sigma)\,[a_1,a_2],\quad \sigma\in
S_2.
\end{equation}
It turns out that the ternary $p$-commutator has the similar properties
with respect to the group $S_3$. It is well known that the group $S_3$ has
two conjugate representations by cubic roots of unit and complex
conjugation. Replacing the ordinary complex conjugation by the conjugation
(\ref{con-of-commutator}) and denoting the above mentioned representations
by $s,{\bar s}$ we get
\begin{eqnarray*}
\lefteqn [\, f_{\sigma(1)},f_{\sigma(2)},f_{\sigma(3)} ]_{\bar q}&=&
      s(\sigma)([ f_1,f_2,f_3 ]_{\bar q}),\cr
[ f_{\sigma(1)},f_{\sigma(2)},f_{\sigma(3)} ]_{q}&=&
      \bar s(\sigma)([ f_1,f_2,f_3 ]_{q}),
\end{eqnarray*}
where $\sigma\in S_3$. 
We have denoted by $s$ the representation of $S_3$ such that
$s(\tau)=q, s(\chi)=\dagger$, where
$$
\tau=\left(\matrix{1 & 2 & 3\cr
                   2 & 3 & 1\cr}\right),\quad
\chi=\left(\matrix{1 & 2 & 3\cr
                   3 & 2 & 1\cr}\right).
$$
Then the conjugate representation ${\bar s}$
takes on the form $\bar s(\tau)={q}^2, \bar s(\chi)=\dagger$.

\vskip .3cm
\noi
{\bf 2. The structure of the algebra ${\bf {\cal T}_q^0(3)}$.} In this
section our 
main concern is the structure of the vector space ${\cal T}_q^0(3)/\{1\}$
with respect to ternary commutator (\ref{p-commutator}) with $p=q$. Let us
remind that if $Cl(n)$ is the classical Clifford algebra generated by
$\gamma_1,\gamma_2,\ldots,\gamma_n$ then the vector space of its even
elements of the form $\varsigma(\omega)=1/4\,\gamma^t\,\omega\,\gamma$, where
$\omega\in so(n,{\bf C})$ is a skew-symmetric $n\times n$-matrix, is the Lie
algebra. Moreover there is the identity
\begin{equation}
[\varsigma(\omega),\varsigma(\omega')]=
          \varsigma([\omega,\omega']),\qquad \omega,\omega'\in so(n,{\bf C})
\label{classical-identity}
\end{equation}
which clearly shows that $\varsigma:so(n,{\bf C})\to Cl^0(n)$ is the
homomorphism of the Lie algebras.  

If $N=1$ then the vector space $T_q^0(N)/\{1\}$ consists only of zero
vector. If $N=2$ then the same vector space is spanned by the monomials
$Q_{\bar 1}\,Q_{2},\;Q_{1}\,Q_{\bar 2}$. Since these monomials commute
with each other the space $T_q^0(2)/\{1\}$ is
trivial with respect to ternary commutator.

The case $N=3$ provides us with the first example of a non-trivial
structure with respect to ternary commutator. In this case the dimension
of the vector space underlying the algebra ${\cal T}_q(N)$ is $3^N=27$. The
vector space $T^0_q(N)/\{1\}$ has dimension 8 and it is spanned by the
monomials $\Sigma_{\mu},\Sigma_{\bar \mu}, \mu=0,1,2,3$ expressed in terms
of generators as follows
$$
\begin{array}{ll}
\Sigma_{\bar 0}=Q_1Q_2Q_3&\quad\quad 
           \Sigma_0=Q_{\bar 1}Q_{\bar 2}Q_{\bar 3}\\
\Sigma_{\bar 1}=Q_{\bar 1}Q_2&\quad\quad
           \Sigma_1=Q_{1}Q_{\bar 2}\\
\Sigma_{\bar 2}=Q_{\bar 2}Q_3&\quad\quad
                   \Sigma_{2}=Q_{2}Q_{\bar 3}\\
\Sigma_{\bar 3}=Q_{\bar 1}Q_3&\quad\quad
                \Sigma_3=Q_{1}Q_{\bar 3 }
\end{array}
$$
Each of these monomials can be written in the form 
$Q_1^{\alpha_1}Q_2^{\alpha_2}Q_3^{\alpha_3}$, where $\alpha_i, i=1,2,3$ 
take the values $0,1,2$, at least one of them is different from zero and
$\alpha_1+\alpha_2+\alpha_3\equiv 0\,(\mbox{mod}\,3)$. Then the ternary
commutator of any three monomials is expressed in terms of the same
monomials as follows
\begin{equation}
[Q^{(\alpha)},Q^{(\beta)},Q^{(\gamma)}]=
C_{\alpha\beta\gamma}\;Q_1^{\delta_1}Q_2^{\delta_2}Q_3^{\delta_3}, 
\label{structure-equations}
\end{equation}
where $Q^{(\alpha)}=Q_1^{\alpha_1}Q_2^{\alpha_2}Q_3^{\alpha_3}$,
$\delta_i=\alpha_i+\beta_i+\gamma_i(\mbox{mod}\,3)$ and the structure
constants 
\begin{eqnarray*}
\lefteqn C\,\;_{\;\alpha\beta\gamma}=
      {\bar q}^l\,(q^{m(\alpha,\beta,\gamma)}&+ &
  q\,q^{m(\beta,\gamma,\alpha)}+{\bar q}\,q^{m(\gamma,\alpha,\beta)}
     +\;\\ & & q^{m(\gamma,\beta,\alpha)}+
 q\,q^{m(\alpha,\gamma,\beta)}+
            {\bar q}\,q^{m(\beta,\alpha,\gamma)}),
\end{eqnarray*}
where $l=\alpha_3\beta_3+\alpha_3\gamma_3+\beta_3\gamma_3,\; 
m(\alpha,\beta,\gamma)=\alpha_2\beta_1+\alpha_2\gamma_1+\beta_2\gamma_1.$
The formula (\ref{structure-equations}) proves that {\it the space
$T_q^0(3)$ is closed with respect to the ternary commutator}. Indeed, the
grade of the monomial at the right-hand side of
(\ref{structure-equations}) is
$\delta_1+\delta_2+\delta_3\equiv 0\,(\mbox{mod}\,3)$. Moreover, 
the monomial at the right-hand side of (\ref{structure-equations}) can not
be equal to the identity element of the algebra since if
$\delta_1=\delta_2=\delta_3=0$ then
$$
m(\alpha,\beta,\gamma)= 
  m(\beta,\gamma,\alpha)=m(\gamma,\alpha,\beta),\;
     m(\gamma,\beta,\alpha)=m(\alpha,\gamma,\beta)=
            m(\beta,\alpha,\gamma)
$$
and these relations imply that $C_{\alpha\beta\gamma}=0$.

\vskip .3cm
\noi
{\bf 3. Representation by cubic matrices.} The fact proved in the previous
section that the space $T_q^0(3)$ is closed with respect to the
ternary commutator can be used to construct the ternary commutator of
cubic matrices. In analogy with the classical identity
(\ref{classical-identity}) we define the ternary commutator of cubic
matrices by the formula
\begin{equation}
\zeta([R^{1},R^{2},R^{3}])=
          [\zeta(R^{1}),\zeta(R^{2}),\zeta(R^{3})],
\label{cm-commutator}
\end{equation}
where $R^{a}, a=1,2,3$ are cubic matrices whose entries we denote
respectively by 
$\rho^{a}_{ABC}$ and $\zeta(R^{a})$ are
the elements of $\three$ expressed in terms of generators as follows
\begin{equation}
\zeta(R^{a})={1\over 6}\,\rho^{a}_{ABC}\,Q_AQ_BQ_C.
\label{form}
\end{equation}
In order to get the linear combination of the monomials
$\Sigma_{\mu},\Sigma_{\bar \mu}$ at
the right-hand side of the above formula we impose the following
requirements on the entries of cubic matrices $R^{a}$ : {\it i)}
$\rho^{a}_{ABC}=0$ if the triple $\{A,B,C\}$ contains both the elements
of the set $\{1,2,3\}$ and the set $\{{\bar 1},{\bar 2},{\bar 3}\}$; {\it
ii)} 
$\rho^{a}_{ABC}={\bar q}_{AB}\,\rho^{a}_{BAC},\;\rho^{a}_{ABC}=
{\bar q}_{BC}\,\rho^{a}_{ACB}$; {\it iii)} $\rho^{a}_{kkl}=\rho^{a}_{{\bar
k}{\bar l}{\bar l}}=\rho^{a}_{{\bar k}l}$ for each pair of indices
$(k,l)$. These requirements lead to the relation
$$
\sum_{\sigma\in S_3} \rho^a_{\sigma(A)\sigma(B)\sigma(C)}=0,
$$
which shows that the cubic matrices $R^a$ can be viewed as a
cubic analogues of skew-symmetric square matrices.

In order to derive the explicit formula for ternary commutator of cubic
matrices we write the element $\zeta(R^{a})$ (\ref{form}) by
means of a slightly different notations. The independent entries of the
cubic matrix $R^{a}$ can be written in the form 
$\rho^{a}_{A({\alpha_1},{\alpha_2},{\alpha_3})}$, where
$A({\alpha_1},{\alpha_2},{\alpha_3})=(A_{\alpha_1},A_{\alpha_2},A_{\alpha_3}),
\alpha_i=0,1,2,\; \alpha_1+\alpha_2+\alpha_3\equiv 0\,(\mbox{mod}\,3)$
and 
\begin{equation}
A_{\alpha_i}=\left\{\begin{array}{ll}
  {\hat i},   & \mbox{$\alpha_i=0$}\\
  i,   & \mbox{$\alpha_i=1$}\\
  {\bar i}, & \mbox{$\alpha_i=2$}
 \end{array}
\right.
\label{A-definition}
\end{equation}
where the hat over the index means that it is omitted. For instant 
$\rho^{a}_{A({0},{1},{2})}=\rho^{a}_{2\bar 3}$. Then the ternary
commutator of cubic matrices is the cubic matrix whose entries are expressed
as follows 
\begin{equation}
[R^1,R^2,R^3]_{D({\delta_1},{\delta_2},{\delta_3})}=
         \sum_{\alpha,\beta,\gamma}\,C_{\alpha\beta\gamma}\;
               \rho^1_{A({\alpha_1},{\alpha_2},{\alpha_3})}
                   \rho^2_{B({\beta_1},{\beta_2},{\beta_3})}
                      \rho^3_{C({\gamma_1},{\gamma_2},{\gamma_3})},
\label{cubic-matrices}
\end{equation}
and the sum at the right-hand side is taken over all triple
$\alpha,\beta,\gamma$ such that $\alpha_i+\beta_i+\gamma_i\equiv \delta_i$.
Obviously the above formula gives only the independent entries of the
cubic matrix $[R^1,R^2,R^3]$. All others entries can be found by means of
the requirements {\it i)-iii)}. We expect that the ternary commutator
(\ref{cubic-matrices}) written in the terms of the entries $\rho_{ABC}^a$
will induce the ternary multiplication of cubic matrices $R^a$.

\section{Applications: generalization of the algebra of supersymmetries}

{\bf 1. Ternary ${\bf Z_3}$-graded commutator}.  It is well known that the
classical Clifford algebras can be used to construct the Lie
superalgebras. The main tool of the construction is the notion of
the $Z_2$-graded commutator which includes both the ordinary commutator
and anticommutator. The simplest example is the classical Clifford algebra
with one generator $\gamma$ such that $\gamma^2=-1$. This algebra is the
Lie superalgebra with even part consisting of the identity element $1$ and
the odd part consisting of the generator $\gamma$. The commutation
relations of this Lie superalgebra have the form
\begin{equation}
\lbrack 1,1 \rbrack_{Z_2}= \lbrack 1,\gamma \rbrack_{Z_2}=0,\quad
	\lbrack \gamma,\gamma \rbrack_{Z_2}=-2.
\label{Z2-supersymmetries}
\end{equation}

We shall use the ternary Clifford algebras to construct the $Z_3$-graded
generalizations of Lie superalgebras. The main tool to be used is the
notion of the $Z_3$-graded ternary $p$-commutator which we define by the
formula 
\begin{equation}
[f_1,f_2,f_3]_{p,{Z_3}}=f_1*f_2*f_3+p^{\pi(a,b,c)}\;f_2*f_3*f_1+
    {\bar p}^{\pi(a,b,c)}\;f_3*f_1*f_2,
\label{graded-commutator}
\end{equation}
where $f_1,f_2,f_3$ are elements of $\TCA$ whose gradings respectively are
$a,b,c$ and $\pi(a,b,c)=abc(a+b)(b+c)(a+c)$ and as before
$p^3=1,\;p\not=1$. Throughout what follows we shall fix $p=q$ in
(\ref{graded-commutator}) omitting the symbol $p$ by the brackets. Let us
consider the TCA with one generator $Q$ 
such that $Q^3=-1$. We define its $Z_3$-grading associating grade 0 to the
identity element and grade 2 to its generator $Q$. Then this TCA provides
us with the simplest example of what may be called the $Z_3$-graded
generalization of Lie superalgebra. It can be easily verified that the
commutation relations of this algebra have form
\begin{eqnarray}
[ Q,Q,Q ]_{Z_3}=-6&\quad
    [ Q^2,Q^2,Q^2 ]_{Z_3}=-6\cr
[ Q,Q,Q^2 ]_{Z_3}=0&\quad
   [ Q,Q^2,Q^2 ]_{Z_3}=0
\label{Z3-graded-algebra}
\end{eqnarray}
and all commutators containing the identity element are equal to zero.

\vskip .3cm
{\bf 2. ${\bf Z_3}$-graded generalization of the algebra of
supersymmetries}. We begin this section by reminding how one can construct
the simplest algebra of supersymmetries making use of the Lie superalgebra
(\ref{Z2-supersymmetries}). This can be done by replacing the identity
element 1 by the operator $P=i\,\partial_t$ acting on the one dimensional
space with the real coordinate $t$ and the generator $\gamma$ by the
operator $S=\partial_{\xi}-i\,\xi\,\partial_t$, where $\xi$ is the
anticommuting coordinate or the generator of the Grassmann algebra. Then 
one obtains the simplest algebra of supersymmetries
\begin{equation}
\lbrack P,P \rbrack_{Z_2}= \lbrack P,S \rbrack_{Z_2}=0\qquad
  \lbrack S,S \rbrack_{Z_2}=-2P.
\end{equation}

In order to construct the $Z_3$-graded generalization of the above algebra
by means of the algebra (\ref{Z3-graded-algebra}) we have to realize the
generator $Q$ by the operator acting on the ternary analogue of the
classical Grassmann algebra. This analogue with one generator is easily
constructed and it is an associative algebra over ${\bf C}$ generated by
$\theta$ such that $\theta^3=0$. There are few different ways to construct
the $N$-extended version of this algebra and they can be found in 
\cite{Abramov},\cite{Abramov-Kerner-Roy}. The derivative with respect to
the generator $\theta$ is defined as follows
\begin{equation}
\tder (1)=0,\qquad\tder (\theta)=1,\qquad\tder (\theta^2)=-q\theta.
\end{equation}
Then the generator $Q$ can be represented by the operator
$q\,\partial_{\theta}+\theta^2$. Finally defining the operators
\begin{equation}
{\cal P}=\partial_t\qquad {\cal Q}=q\,\partial+\theta^2\partial_t,
\end{equation}
we get the algebra
\begin{eqnarray}
\lbrack {\cal Q},{\cal Q},{\cal Q}\rbrack_{Z_3}
		  =-6 {\cal P}&\quad
\lbrack {\cal Q}^2,{\cal Q}^2,{\cal Q}^2\rbrack_{Z_3}
		  =-6 {\cal P}^2\cr
\lbrack {\cal Q},{\cal Q},{\cal Q}^2\rbrack_{Z_3}=0&\quad
\lbrack {\cal Q},{\cal Q}^2,{\cal Q}^2\rbrack_{Z_3}=0.
\end{eqnarray}
The commutators containing either the operator ${\cal P}$ or ${\cal P}^2$
are equal to zero. The remarkable peculiarity of this algebra is that it
contains the second derivative with respect time. 
%%%%%%%%%%%%%%%%%%%%%%%%%%%%%%%%%%%%%%%%%%%%%%%%%%%%%%%%%%%%%%%%%%%%%%%%
%@@@
\vskip .4cm
\centerline{\bf Acknowledgements}
\vskip .4cm
\indent
I am grateful to the organizers of the XXI International Colloquium
on Group Theoretical Methods in Physics for financial assistance and
hospitality during the 
Colloquium. I thank Raul Roomeldi for discussions and computer 
program helping to calculate the ternary commutators.
This work was supported in part by the Estonian Science Foundation 
under the grant No. 2403.
%%%%%%%%%%%%%%%%%%  REFERENCES %%%%%%%%%%%%%%%%%%%%%%%%%%%%%%%%%%%%%%%%


\begin{thebibliography}{10}

\bibitem{Abramov}
V. Abramov,  {\em Ternary generalizations of Grassmann algebra,}
Proc. Estonian Acad. Sci. Phys. Math., {\bf 45}, 2/3, 174-182, 1996. 
\bibitem{Abramov-Kerner-Roy}
V. Abramov, R. Kerner, B. Le Roy, {\em Hypersymmetry: a $Z_3$-graded
generalization of Supersymmetry}, J.Math.Phys., (to appear).
\bibitem{Colatto}
L.P. Colatto and J.L. Matheus-Valle, {\em On $q$-deformed supersymmetric
classical mechanical models}, IC/95/96.
\bibitem{Macfarlane}
J.A. de Azc\'arraga and A.J. Macfarlane, {\em Group theoretical
foundations of fractional supersymmetry}, FTUV/95-23, IFIC/95-23.
\bibitem{Mohammedi}
N. Mohammedi, {\em Fractional Supersymmetry}, hep-th/9412133.
\bibitem{Trautman}
A. Trautman, {\em Spinors and the Dirac operator on hypersurfaces. I:
General theory}, J.Math.Phys. {\bf 33} (12), 4011-4019, 1992.

\end{thebibliography}
\end{document}